\documentclass[twocolumn,amsmath,amssymb,prl]{revtex4}

\usepackage{graphicx}
\usepackage{color}
\usepackage{hyperref} 

\begin{document}

\title{Observation of a gradient catastrophe generating solitons}
\author{Claudio Conti,$^{1}$ Andrea Fratalocchi,$^{2,1}$ Marco Peccianti,$^{3,1}$ Giancarlo Ruocco,$^{1,4}$ Stefano Trillo$^{1,5}$}
\email{claudio.conti@roma1.infn.it}
\affiliation{
$^{1}$Research center SOFT INFM-CNR Universit\`{a} di Roma ``La Sapienza'',  P. A. Moro 
2, 00185, Roma, Italy\\
$^{2}$Centro Studi e Ricerche ``Enrico Fermi'', Via Panisperna 89/A, 
00184 Rome, Italy \\
$^{3}$ INRS-EMT University of Quebec, 1650 Blvd. Lionel Boulet,  Varennes, Quebec,Canada\\
$^{4}$ Dipartimento di Fisica, Universit\`{a} di Roma ``La Sapienza'',  P. A. Moro 
2, 00185, Roma, Italy\\
$^{5}$ Dipartimento di Ingegneria, Universit\`{a} di Ferrara, Via Saragat 1, 
44100 Ferrara, Italy
}
\date{\today}  
\begin{abstract}
We investigate the propagation of a dark beam in a defocusing medium in the strong nonlinear regime. 
We observe for the first time a shock fan filled with non-interacting one-dimensional grey solitons that 
emanates from a gradient catastrophe developing around a null of the optical intensity.  
Remarkably this scenario turns out to be very robust, persisting also when the material nonlocal response 
averages the nonlinearity over dimensions much larger than the emerging soliton filaments.
\end{abstract}

\maketitle
{\em Introduction} 
In many physical systems propagation phenomena are affected primarily 
by the interplay of dispersive and nonlinear effects. In this context,
solitons (or solitary waves), i.e. wave-packets that stem from a mutual balance between the two effects,
account successfully for several phenomena ranging from long-span non-spreading propagation 
and elastic interactions of beams, to the coherent behavior of ensembles of particles, e.g. ultracold atoms, or coupled oscillators. 
Studies in this field were mainly focused on individual solitons or interactions between 
them. However, several solitons can emerge at once from breaking of large amplitude smooth waves \cite{Zabusky65}, 
as for instance observed in oceanography \cite{Smith88}. While theoretical studies indicates
the phenomenon to be generic \cite{Kamchatnov02,Bett06}, the observation of such 
multi-soliton regime in reproducible lab experiments has been elusive. 
\newline \indent
In this letter, we report a lab experiment in optics which demonstrates that a fan of non-colliding 1D solitons emerge, 
owing to a gradient catastrophe (i.e. an infinite gradient developing from a smooth input) 
developing around a zero of the field. 
Specifically we consider a dark-like optical beam 
(i.e., a dark stripe on a bright background) and operate, unlike previous experiments on dark solitons 
\cite{Kivshar92,Chen97}, in a regime where nonlinearity outweighs diffraction 
(i.e., power of the background largely exceeding that needed to trap a fundamental dark soliton).
In this regime, we are able to monitor directly the evolution along a thermal defocusing medium.
We observe the formation of a dark focus point which corresponds to a gradient catastrophe 
of the hydrodynamic type around a point of vanishing intensity. 
The infinite gradient of the hydrodynamic stage is regularized by the presence of weak diffraction, 
which causes the appearance of fast oscillations in an expanding region (fan),
a feature common to the wide class of so-called collisionless or dispersive shock waves (DSW) or undular bores,
investigated theoretically in several contexts
\cite{Gurevich74,Bronski94,Kodama95,Forest98,Forest99,Porter02,Damski04,PerezGarcia04,El05,Fratax08}.
\newline \indent
In our setting, the DSW is essentially composed by 1D dark soliton filaments, 
which become manifest after the catastrophe point, 
and maintain fixed parameters (velocity and darkness) as soon as they emerge \cite{Fratax08}, also
not exhibiting the rapid decay into vortices characteristic of shock waves in superfluids \cite{Dutton01}.
Our scenario turns out to be remarkably robust against the nonlocal character of the nonlinearity,
and presents also significant differences with DSW resulting from bright disturbances
\cite{Grischkowsky89,Hoefer06,Wan07,Gofra07}. In the latter case
the catastrophe occurs, indeed, at finite transverse extension, giving rise, in 1+1D, to two symmetric fans 
(connected by a quasi-flat background) \cite{Grischkowsky89,Wan07}, 
while the relative oscillations change dynamically 
(i.e., dark solitons in the train have always slowly-varying parameters upon propagation).
\begin{figure}[h]
\includegraphics[width=8.3cm]{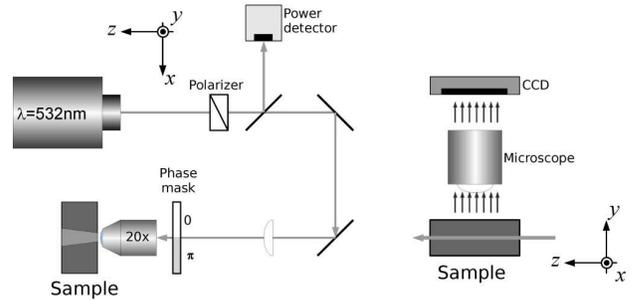}
\caption{\label{setup} (Color online) Sketch of the experimental setup.}
\end{figure}

{\em Experiment} Our sample consists of a BK7 glass cell of dimensions 1cm x 4cm x 1cm in the $X$ $Y$ and $Z$ (propagation) 
direction respectively, containing a solution (concentration $C=0.055 mMol/dm^3$) of Rodhamine-B in Methanol. 
A beam at $\lambda_0=532$ nm from a diode-pumped CW Nd-Vanadate laser was focused down to a strongly elliptical beam
(ellipticity 1:30) by means of a cylindrical lens of focal length $L_f=100 mm$ and a 20X microscope objective. The beam is coupled 
in the cell at $Z=0$ lying in the cell mid-plane (sufficiently far from bottom, top, and lateral liquid-glass interfaces to avoid
the dynamics to be significantly affected by boundary conditions). 
A phase mask is placed on the beam path resulting in an abrupt change of $\pi$ in the optical phase across the line $X=0$. 
After the mask we let the beam diffract shortly and then we focus it onto the sample, 
producing an input background bright beam of dimension $600 \times 20$ $\mu$m, 
onto which a dark stripe (with zero intensity in $X=0$ and hyperbolic-tangent-like $X$-profile) is nested.
The stripe is parallel to the narrower spot-size ($Y$ direction), while along $X$
the dark notch of $25  \mu m$ FWHM sees a quasi-constant background due to the $600 \mu$m width.
We detect no significant changes along $Y$ over the propagation lengths involved,
witnessing that our arrangement mimics a strict 1+1D ($X-Z$) setting.

\begin{figure}[h!]
\includegraphics[width=9cm]{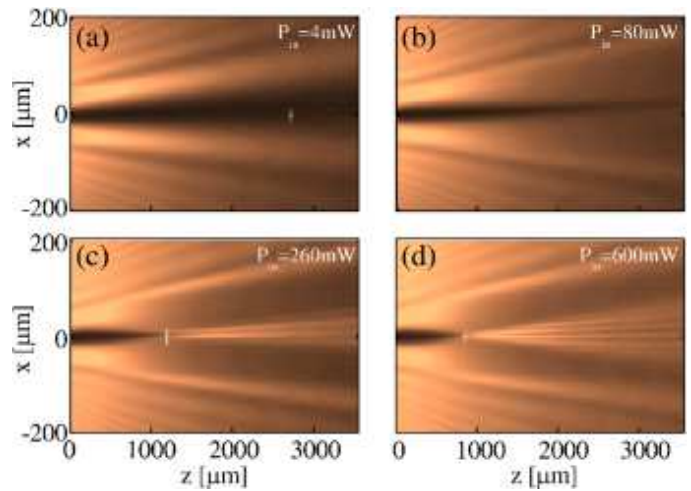} %
\caption{\label{collection} 
(Color online) Transverse distribution of the intensity along the cell  ($x-z$ plane),
as observed from top scattered light for four different input powers.
Superimposed light curves (yellow) are retrieved intensity profiles at $Z=0.6$ mm,
while in (d) the right dark (blue) curve is relative to $Z=2.25$ mm.
The diffraction fringes aside the central dark notch are due to the focusing
system before the sample and reflect the slight convex wavefront of the
bright background.
}
\end{figure}

The power coupled into the sample is measured by means of a beam splitter in the front of the laser output and a silicon detector.
As sketched in Fig. \ref{setup} a microscope and a charge coupled device (CCD) camera allows us to collect the light scattered in the vertical (Y) direction above the cell, so to produce a direct planar image of the relevant beam evolution ($X-Z$ plane).
In Fig. \ref{collection} we show the field intensity distribution collected in this way at different input powers $P_{in}$.

\begin{figure}[h]
\includegraphics[width=8.7cm]{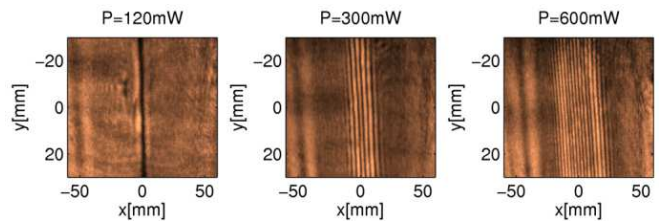}
\caption{\label{farfield} 
(Color online) Far-field intensity distribution in the transverse plane $X-Y$ 
at the sample output for different input powers:
the intensity is collected after about 1 meter 
of free-air propagation.}
\end{figure}

At low power ($4 mW$) the dark notch diffracts, broadening in propagation toward positive $Z$,
whereas at higher power the nonlinear thermal response of the sample counteracts the diffraction 
leading to exact counterbalance (dark soliton formation, $P_{in} \simeq 80mW$), 
and subsequent overall focusing.  
By further increasing the power to $P_{in}=260mW$ (Fig. \ref{collection}c) and $P_{in}=600mW$ (Fig. \ref{collection}d),  
the dark notch undergoes to a clear focus point (gradient catastrophe or breaking point).
Beyond such point, the beam undergoes non-trivial breaking forming a DSW constituted by
narrow dark soliton filaments which fill progressively a characteristic fan.
The number of dark filaments in the fan grows with power. This is also clear from the measured far-field 
relative to the output of our cell (i.e., after $Z=1$ cm of propagation) displayed in Fig. \ref{farfield}. 
Importantly, these data clearly shows that the filaments, once seen in the transverse plane, 
maintain the stripe features of the input (similarly to the bright case \cite{Wan07}), 
not exhibiting any transverse instability
or decay into vortices characteristic of other superfluidity and optical experiments 
\cite{Dutton01,Mamaev96,Kivshar92}, 
allowing for a description in terms of pure 1+1D ($X-Z$) dynamics.

{\em Theory} 
The dynamics observed experimentally can be studied and understood on the basis
of the following generalized nonlinear Schr\"{o}dinger (NLS) model
\begin{eqnarray}
\displaystyle i  \varepsilon \frac{\partial \psi}{\partial z} +
\frac{ \varepsilon^2}{2} \frac{\partial^2 \psi}{\partial x^2} - \delta n \psi=-i \frac{\alpha}{2} \varepsilon \psi, \label{nls1} \\
-\sigma^2  \frac{\partial^2 \delta n}{\partial x^2} +  \delta n = |\psi|^2, \label{nls2}
\end{eqnarray}
where the first equation stands for the paraxial nonlinear wave equation
for $\psi \equiv A/\sqrt{I_0}$, i.e. the beam envelope $A=A(Z,X)$ normalized to peak intensity $I_0$
(in the experiment $I_0=P_{in}/A_e$, where $A_e$ is the background beam area).
The transverse and longitudinal coordinates $x,z=X/w_0,Z/L$ are scaled to the waist of the input dark notch $w_{0}$, 
and the geometric mean $L \equiv \sqrt{L_{nl} L_d}$ between the length scales 
$L_{nl}=(k_{0} |n_{2}| I_0)^{-1}$ and $L_d= k  w_0^2$  characteristic of the nonlinear
and diffractive terms, respectively ($n_{2}$ is the Kerr coefficient that characterizes
an index change of the local type $\Delta n=n_{2} |A|^{2}$), and
$\alpha=\alpha_{0} L$ is the normalized attenuation constant. Such scaling 
allows us to highlight the fact that we operate in the weakly dispersive case, 
where the model mimics the quantum Schr\"{o}dinger equation with the smallness 
parameter $\varepsilon \equiv L_{nl}/L=\sqrt{L_{nl}/L_d}$ playing the role of Planck constant.
The normalized refractive index change $\delta n=k_{0} L_{nl} \Delta n$ acts as a 
self-induced potential driven by the normalized intensity profile $|\psi(x)|^2$.
The free parameter $\sigma^2$ measures the diffusion length and gives
the degree of nonlocality of the nonlinear response.
This model describes the nonlocal features of the nonlinear response
with sufficient accuracy regardless of their physical origin.
Specifically, the model was shown to give an excellent description of thermal nonlinearity \cite{Gofra07,Yakimenko05},
while it allows for a reduction to the integrable semiclassical 
NLS equation in the local and lossless limit $\sigma^2=\alpha=0$.

\begin{figure}
\includegraphics[width=9cm]{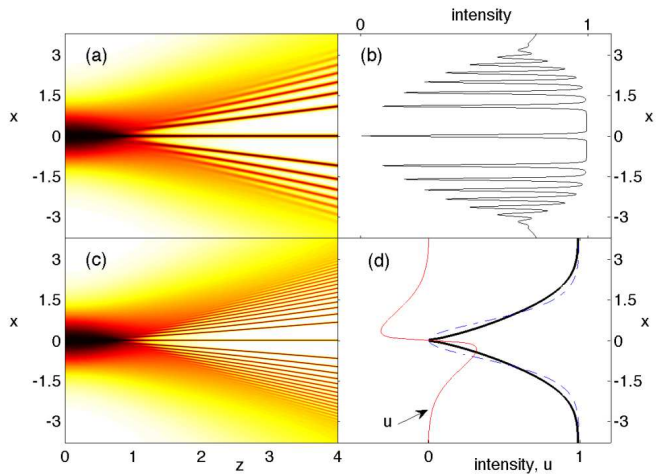}
\caption{(Color online) Numerical simulations of Eqs. (1-2) with $\sigma=\alpha=0$
and $\psi_0=tanh(x)$:
(a-b) Level plots of $|\psi|^2$ for $\varepsilon=0.05$
(a) and $\varepsilon=0.02$ (c);
(b) snapshot of case (a) at $z=4$; 
(d) snapshot of case (c) at the catastrophe point $z=0.75$:
frequency $u$ (solid red curve and intensity (black solid curve) 
compared with the input (blue dashed curve).
\label{theory1}}
\end{figure}

\begin{figure}
\includegraphics[width=9cm]{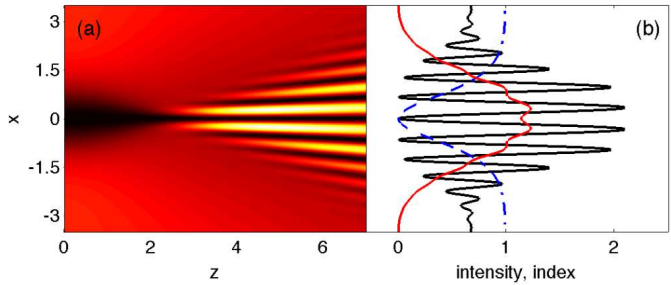}
\caption{(Color online) 
Evolution in the nonlocal and lossy case: (a) Level plot of intensity
(the color scale is adapted to compensate for losses); (b) snapshots at $z=7$,
of intensity (black solid curve) and relative index change $\delta n$ (red solid curve). 
The input is the dashed blue curve.
Here $\varepsilon=0.05$ ($P \simeq 0.6 W$ in our experiment), $\sigma^2=1$, $\alpha=0.3$. 
The corresponding physical breaking (catastrophe) distance turns out to be $Z_b=L z_b=0.8 mm$.
\label{theory2}}
\end{figure}

The essential physics can be explained indeed by the latter limit, for which
the outcome of numerical computations based on Eqs. (1-2) with input $\psi_0(x)=\tanh(x)$,
are displayed in Fig. \ref{theory1} for two different values of $\varepsilon$ (powers). 
The initially dominant nonlinearity allows us to 
adopt a description in terms of hydrodynamical variables $\rho$
and $u \equiv \partial_x S$. This is made by applying the WKB transformation 
$\psi(x,z)=\sqrt{\rho(x,z)} \exp \left[i S(x,z) \right/\varepsilon]$
\cite{Kamchatnov02,Bronski94,Kodama95,Forest98,Hoefer06},
which allows to reduce Eqs. (1-2), at lowest order in $\varepsilon$, to the following 
system written in the form of hyperbolic conservation laws \begin{equation} \label{hyp}
\frac{\partial {\bf a}}{\partial z} + \frac{\partial {\bf f}}{\partial x}=0;~{\bf a} \equiv
\left( \begin{array}{c}
  \rho   \\ q   \\  
\end{array}\right);~{\bf f}({\bf a}) \equiv
\left( \begin{array}{c}
  q   \\ \frac{\rho^{2}}{2}+ \frac{q^{2}}{\rho} \\  
\end{array} \right),
\end{equation}
where $q(z,x) \equiv \rho(z,x) u(z,x)$.
Equation (\ref{hyp}), that rules classical 1D dynamics of an isentropic gas or shallow water ($u$ being,
in this case the velocity of the gas or water, and $\rho$ the gas density or the water level), 
predicts that the dynamics of the input hole in the density $\rho(x,0)=|\psi_0|^2$ produces a gradient of "velocity" $u$, 
whose sign ($u$ turns out to be positive for $x<0$, and viceversa)
is such to give rise to compressional waves. 
Equivalently, due to the defocusing nature of the medium
the central dark region has a higher index which draws light inwards.
As a result the input dark notch experiences
a dramatic steepening and focusing around its null, which in turn enforces the velocity gradient, 
until eventually a singularity (gradient catastrophe) develops at a finite distance, consistently
with the hyperbolic nature of Eqs. (\ref{hyp}). 
Such singularity is characterized by crossing of characteristics
associated with Eqs. (\ref{hyp}) and become manifest as a vertical front in the variable $u$
and a cusp in the intensity $\rho$, as displayed in Fig. \ref{theory1}d (numerical
results from Eqs. (\ref{hyp}) are exactly coincident to those shown).
However, when such a high (virtually infinite) gradients develop, the hydrodynamic
description breaks down, and diffraction regularizes the front through the appearance 
of an expanding region of fast oscillations, i.e. a shock fan. 
The shock fan is progressively filled with non-interacting dark filaments,
whose angle (transverse velocity) increases as the relative darkness decreases,
which is a universal feature of dark (grey) solitons \cite{Kivshar92}. 
Furthermore the number of filaments (solitons) increases 
for smaller $\varepsilon$ (higher powers). In particular, 
for $1/\epsilon=N$, $N$ integer, one can show that
the asymptotic state is made by $2N-1$ dark solitons, while no radiation at all is produced \cite{Fratax08}. 
In other words, the solitons which are embedded in the input
become manifest at sufficiently large distance, though this occurs in a non trivial way through
a critical behavior characterized by a cooperative initial stage which results into the catastrophe focus point.
Our experiment provides evidence that nonlocality and losses do not qualitatively affect this scenario
(at variance with observations for bright Gaussian input, which involves a lot of radiation instead of solitons \cite{Gofra07}).
This is ultimately related to the persistence of solitons in the presence of terms that break the integrability.
We recall that, in our system, the losses and nonlocality are intimately related because
the index change is due to heating caused by the strong absorption of the dye,
while the nonlocality arises from the intrinsic tendency of heat to diffuse \cite{Gofra07}.
As shown in Fig. \ref{theory2}, where we report simulations based on Eqs. (\ref{nls1}-\ref{nls2})
with $\sigma^2=1$, the overall dynamics is quite similar to the local one, 
except for a slight adiabatic broadening of the solitons due to the losses.
The post-catastrophe breaking still occurs in the form of narrow soliton filaments, which 
model (\ref{nls1}-\ref{nls2}) support also in the nonlocal case \cite{Kartashov07}.
This occurs despite the fact that the induced potential $\delta n$ that trap them
becomes, owing to diffusion, a smooth function (see Fig. \ref{theory2}b, red curve) that
no longer follows the deep oscillations of the intensity as in the local case.
Simulations in transverse 2D with elliptical input reminiscent of the experiment confirm
the validity of this 1D picture, allowing us to conjecture that nonlocality stabilizes
the soliton stripes in the fan against transverse instabilities.

\begin{figure}
\includegraphics[width=8.3cm]{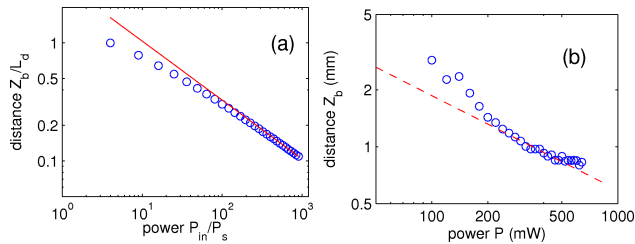}
\caption{ (Color online) Breaking distance $Z_b$ vs. input power $P_{in}$ (log-log scale):
(a) local case, numerical results (dots) vs. behavior $Z_b \propto P_{in}^{-1/2}$ 
(solid line), characteristic of the hydrodynamic limit;
(b) nonlocal case: experimental results (dots) vs. data extrapolated from
numerical simulations (dashed line) performed different values of powers $P_{in}$ ($\varepsilon$).
Here $\sigma=0.3$ was fixed to find the best agreement with the data. 
\label{theoryvsexp}}
\end{figure}

Importantly, the breaking distance 
(point in $Z$ of maximum intensity gradient)
turns out to be significantly affected by the attenuation and the degree of nonlocality,
an issue that we have investigated in detail.
In the local and loss-less case, the hydrodynamic limit yields a constant normalized breaking distance $z_b=0.75$,
corresponding to a physical distance $Z_b=z_b L$ that scales with power as $P_{in}^{-1/2}$.
This is confirmed by numerical solutions of Eqs. (\ref{nls1}-\ref{nls2}) with $\sigma=\alpha=0$,
performed for $\varepsilon=1/N$, $N$ integer (i.e. $P_{in}/P_s=N^2$,  $P_s$ being the fundamental dark soliton power \cite{Kivshar92}).
As shown in Fig. \ref{theoryvsexp}a, $Z_b$ approaches the law $P_{in}^{-1/2}$
for high enough values of $N$. A similar trend is confirmed by numerical simulations
performed in the nonlocal case by employing the parameters of our experiment
(we measured by means of Z-scan apparatus $\alpha_0=1.17$ mm$^{-1}$, $n_2=-2 \times 10^{-10} m^2/W$),
as reported in Fig. \ref{theoryvsexp}b. As shown, the expected breaking distance agrees reasonably well
with the measured data except for very low and high powers, 
the latter showing a saturation effect not accounted for by our model. 
Here we have used $\sigma$ as a free parameter, finding the best agreement
for $\sigma \simeq 0.3$, which is consistent with our independent estimate 
$\sigma=\left[ \frac{D_T \rho_0 c_p |n_2|}{\alpha_0 |dn/dT| w_0^2} \right]^{1/2} \simeq 0.33$ 
\cite{Gofra07}, based on the values of the parameters for methanol
$D_T=10^{-7} m^2/s$, $\rho_0=791 kg/m^3$, $c_p=2 \times 10^3 J kg^{-1}  K^{-1}$, $dn/dT=-4 \times 10^{-4} K^{-1}$.
In summary, we have presented the first demonstration of a gradient catastrophe occurring
around a point of vanishing field in the regime of strong nonlinearity.
The post-breaking dynamics gives rise to a fan of non-interacting 1D soliton filaments.
Due to weak dispersion such filaments are very narrow, yet they are robust against 
nonlocal averaging over much larger widths. 

The research leading to these results has received funding from the European Research
Council under the European Community's Seventh Framework Program (FP7/2007-2013)/ERC grant agreement n. 201766.

\end{document}